\documentclass[12pt,amsmath,amssymb,epsfig]{revtex4}
\usepackage{graphicx}
\usepackage{epsfig}
\usepackage{amssymb}

\def\br{\begin{eqnarray}}
\def\er{\end{eqnarray}}
\def\be{\begin{equation}}
\def\ee{\end{equation}}

\def\({\left(}
\def\){\right)}

\begin{document}

\title{Mass and width of a composite Higgs boson}
  
\author{A.~Doff$^1$ and A.~A.~Natale$^2$ }
\email{agomes@utfpr.edu.br,natale@ift.unesp.br}
\affiliation{$^1$Universidade Tecnol\'ogica Federal do Paran\'a - UTFPR - COMAT
Via do Conhecimento Km 01, 85503-390, Pato Branco - PR, Brazil \\
$^2$Instituto de F\'{\i}sica Te\'orica, UNESP 
Rua Pamplona, 145,
01405-900, S\~ao Paulo - SP,
Brazil}  
                 
\date{\today}

\begin{abstract}
The scalar Higgs boson mass in a Technicolor model  was obtained by Elias and Scadron with the analysis 
of an homogeneous Bethe-Salpeter equation (BSE), however it was performed before the most recent 
developments of walking gauge theories. It was not observed in their work that dynamically generated 
technifermion mass may vary  according to the theory dynamics that forms the scalar bound state. 
This will be done in  this work and we also call attention that their calculation must change to take
into account the normalization condition of the BSE. We compute the width of the composite boson and
 show how the gauge group and fermion content  of a technicolor theory can be inferred from
the measurement of the mass and width of the scalar boson.
\end{abstract}

\maketitle

The fermion and gauge boson masses in the standard model of elementary particles are explained by
their interaction with an elementary Higgs scalar boson. The marvelous
agreement existent between experiment and the standard model theory can be credited to the
electroweak gauge symmetry structure and its breaking through the Higgs mechanism based on this fundamental 
scalar boson. Therefore it is clear that the existence of the Higgs boson is a cornerstone for
the model, and its discovery is a quest for the LHC accelerator. Although the importance of fundamental
scalar bosons in gauge theories is widely accepted it is also true that no one of these bosons
has been found up to now. To this fact we can add that the mechanism of spontaneous symmetry breaking in field
theory and superconductivity involves the existence of composite scalar states~\cite{nl}, and in Quantum
Chromodynamics (QCD) the correspondent scalar boson is considered to be the elusive $\sigma$ meson~\cite{ds,polosa}.
Of course, fundamental scalar bosons are quite natural in the supersymmetric versions of the standard model, but if
supersymmetry is not realized at the Fermi scale, one plausible possibility is that these scalar bosons are composite,
following the ideas of the usually called technicolor theories (TC)~\cite{tc,hs}. 

Masses of fundamental scalars bosons appear in a potential whose couplings are frequently assumed to be perturbative, whereas in the dynamical
symmetry breaking case (or TC theories) the masses are originated from a non-perturbative effective potential~\cite{we},
consequently it is certainly more complicated to infer the values of masses and couplings in this last case.
Precision electroweak measurements indicate a preference for a small Higgs mass~\cite{pdg}, whereas, as we shall
briefly discuss later, the composite scalar boson mass ($m_H$) is usually expected to be of the order of the Fermi scale or the dynamically generated technifermion mass ($m^{TC}$), assumed to be of the order of the TC scale ($\Lambda_{TC}$) as shown by Elias and Scadron~\cite{es}. This is a prejudice originated from QCD where the $\sigma$ meson mass ($m_\sigma$) is of the order of the QCD characteristic scale ($\Lambda_{QCD}$)~\cite{ds}.
However the TC dynamics may be totally different from the QCD one, and $m_H$ may have a more subtle dependence
on the fermion content of the TC theory. Actually, Sannino and collaborators have been claiming that in
walking technicolor theories the composite Higgs boson may be quite light~\cite{san}, and
recently we confirmed their results through the calculation of an effective action for technicolor~\cite{we}. 

The Elias and Scadron~\cite{es} calculation of the composite scalar mass is simple and elegant, based on the similarities of homogeneous Bethe-Salpeter equation (BSE) for scalar states and the Schwinger-Dyson equation for the fermionic self-energy. However it was performed before the most recent developments of walking gauge theories~\cite{walk} and assumed a standard operator product expansion behavior (OPE) for the dynamically generated masses. It was not observed in their work that $m^{TC}$ may vary according the dynamics of the theory that forms the scalar bound state, and the result should be written in terms of known standard model quantities and TC theory gauge group and fermion content~\cite{frere}.  We verified that their calculation must be changed to take into account the normalization condition of the BSE, which is not important if TC is just an scaled QCD version, but it is important in the case of walking gauge theories. We use a quite general expression for the technifermion self-energy that spans the full set of possible behaviors for the dynamically generated fermion masses of the theory forming the composite scalar state~\cite{doff1}. 

Assuming that the composite Higgs sector is identical to the $SU(2)$ linear sigma model of mesons we compute its width into $WW$ and $ZZ$ bosons and compare our results to the ones of the standard model. We show how the gauge group and fermion content of a TC theory can be discovered  through the mass and width measurements of the composite Higgs boson.

The calculation of Ref.\cite{es}, as well as the one of Ref.\cite{ds}, is quite simple and reminiscent from
the earlier Nambu and Jona-Lasinio model~\cite{nl}. It comes out from the following relation
$$
\Sigma (p^2) \approx  \Phi_{BS}^P (p,q)|_{q \rightarrow 0} \approx \Phi_{BS}^S (p,q)|_{q^2 = 4 m_{dyn}^2 }\,\,\, ,
$$ 
where the solution of the fermionic Schwinger-Dyson equation ($\Sigma (p^2)$), that indicates the
generation of a dynamical techniquark(quark) mass ($m_{dyn}$) and chiral symmetry breaking of TC(QCD), is a solution of 
the homogeneous Bethe-Salpeter equation for a massless pseudoscalar bound state ($\Phi_{BS}^P (p,q)|_{q \rightarrow 0}$),
indicating the existence of Goldstone bosons (technipions\-(pions)), and is also a solution of  the 
homogeneous BSE of a scalar p-wave bound state ($\Phi_{BS}^S (p,q)|_{q^2 = 4 m_{dyn}^2 }$). This imply the existence of 
a scalar boson (the $\sigma$ meson in the QCD case) with a mass given by 
\be
m_{H} \,  = \, 2 m_{dyn}^{TC}\,\,\, ,
\label{eq11}
\ee

In the QCD case Delbourgo and Scadron found~\cite{ds} $m_{\sigma} = 2 m_{dyn}^{QCD}$. As the QCD phenomenology tell us
that $m_{dyn}^{QCD}\approx \Lambda_{QCD}$, we have $m_{\sigma} \approx 600$MeV. In TC, following a direct extrapolation
of the QCD dynamics as in Ref.\cite{es}, we would expect $\Lambda_{TC}\approx m_{dyn}^{TC}$ to be of the order of the Fermi scale and 
$m_H \approx {\cal{O}}(1)$TeV.

We will modify the result mentioned above introducing a quite general expression for the fermionic self-energy 
($\Sigma$). As it is believed that dynamically broken gauge theories do not necessarily have the same dynamical
behavior of QCD, as happens, for example, in the case of walking or conformal technicolor gauge theories \cite{hs,san,walk}, we will
work with a technifermion self-energy that interpolates between  the standard OPE result and the extreme walking technicolor behavior, which is the case where the symmetry breaking is dominated by higher order interactions that are relevant at or above the TC scale, leading naturally to a very hard dynamics~\cite{soni,soni2}. Our ansatz for the self-energy is~\cite{doff1,doff2,we}
\be 
\Sigma (p^2) \sim m_{dyn} \left( \frac{ m_{dyn}^2 }{p^2}\right)^{\alpha}\left[1 + b g^2 \ln\left(p^2/\Lambda^2 \right) \right]^{-\gamma\cos (\alpha \pi)}  \,\,\, .
\label{eq12}
\ee	
In the above expression $\Lambda$ is the characteristic scale of mass generation of the theory forming the
composite Higgs boson, these quantities can be identified with the TC scale ($\Lambda_{TC}$), and for simplicity we
assume $\Lambda_{TC}\approx m_{dyn}^{TC} = m_{dyn}$, $g$ is the TC running coupling constant, $b$ is the coefficient of $g^3$ term in the renormalization group $\beta$ function, $\gamma= 3c/16\pi^2 b$, and  $c$ is the quadratic Casimir operator given by 
\[
c = \frac{1}{2}\left[C_{2}(R_{1}) +  C_{2}(R_{2}) - C_{2}(R_{3})\right] \,\,\, ,
\]
where $C_{2}(R_{i})$,  are the Casimir operators for fermions in the representations  $R_{1}$ and  $R_{2}$ that form a composite boson in the representation $R_{3}$. If the fermion condensation happens in the singlet channel and $R_{1}$ and  $R_{2}$ are in the same representation ($R$) we
simply have $c=C_2 (R)$.
The only restriction on the ansatz of Eq.(\ref{eq12}) is $\gamma > 1/2$ \cite{lane2}, which will be recovered in this work and indicates a condition on the composite wave function normalization. Notice that the standard OPE behavior for $\Sigma (p^2)$ is obtained when $\alpha \rightarrow 1$, whereas the extreme walking technicolor solution is obtained when $\alpha \rightarrow 0$.

The scalar boson mass is given by Eq.(\ref{eq11}). However $m_{dyn}$ in Eq.(\ref{eq11}) should be written in terms of measurable
quantities and by group theoretical factors of the strong interaction responsible for forming the composite scalar boson. The way 
this is accomplished follows the work of Ref.~\cite{es}: $m_{dyn}$ will be related to $F_\Pi$ (the  Technipion decay 
constant), and this last one will be related to the vacuum expectation value (VEV) of the Standard Model through
\be
\frac{g_w^2 n_d F_\Pi^2}{4}  = \frac{g_w^2 v^2}{4} = M_W^2 \,\,\, ,
\label{eq33}
\ee
where $n_d$ is the number of Technifermion doublets, $v \sim 246 GeV$ is the Standard Model VEV and $F_\Pi$ is obtained from the Pagels and Stokar relation~\cite{pagels},  
\be 
F^2_{\Pi} = \frac{N_{{}_{TC}}}{4\pi^2}\int\!\!\frac{dp^2p^2}{(p^2 + \Sigma^2(p^2))^2}\!\!\left[\Sigma^2(p^2) - \frac{p^2}{2}\frac{d\Sigma(p^2)}{dp^2}\Sigma(p^2)\right] \,\,\, .
\label{eq34}
\ee 
At this point it is important to remember that the relation between $F_\Pi$ and $m_{dyn}$ will depend strongly on
the $\Sigma (p^2)$ behavior described by Eq.(\ref{eq12}), which is one of the differences that we have with Ref.~\cite{es}.
Similarly to the procedures of Ref.\cite{doff1,doff2,doff3,we} we can determine the values of $m_H$ in the limits $\alpha =0 $ and $\alpha = 1$ which are given by
\br
&& m_H^{(0)} \approx  2\left[v\left(\frac{8 \pi^2  bg^2(2\gamma -1)}{N_{TC}n_F} \right)^{1/2}\right]
\label{eq410} \\
&& m_H^{(1)} \approx 2\left[\sqrt{\frac{4}{3}}v\left( \frac{8\pi^2}{N_{TC}n_{F}}\right)^{1/2}\right].
\label{eq41}
\er 
Where we have used Eqs.(\ref{eq11}), (\ref{eq12}), (\ref{eq33}) and (\ref{eq34}) to obtain expressions that are functions of
$N_{TC}$ (we are considering $SU(N_{TC})$ TC gauge groups), $n_d (\equiv n_F/2)$, $b$, $c$ and $v$.

The expressions for the scalar boson masses appearing above are a guide for the minimum and maximum values of the $\alpha$ parameter
($0$ and $1$) present in Eq.(\ref{eq12}), that correspond respectively to the extreme walking TC behavior and the scaled QCD 
behavior for the TC theory. The full numerical calculation of $m_H$ for arbitrary $\alpha$ values is bounded by these last values.
This variation of the ansatz with $\alpha$ is what makes our calculation a general one, it covers all
possible solutions of the Schwinger-Dyson equation (or Bethe-Salpeter equation) for fermions forming
the composite scalar boson.
In the Table I we show the values  for $m_H$ in the limits $\alpha =0 $ and $\alpha = 1$ obtained with Eqs.(\ref{eq410}) and (\ref{eq41}).

\begin{table}[t]
\begin{center}
\begin{tabular}{ccccc}
 TC group  & $m_H^{(0)}$ &  $n_F^{(0)}$ & $m_H^{(1)}$ &  $n_F^{(1)}$ \\ \\
\hline
   $SU(2)_{TC}$ &  585\,GeV    &   8    &  1.480\,GeV   &    6    \\ 
   $SU(3)_{TC}$ &  414\,GeV &   12     &     1.209\,GeV   &     6    \\ 
   $SU(4)_{TC}$ &  304\,GeV   &   14     &   1.047\,GeV       &   6    \\   
\hline                   
\end{tabular}
\end{center}
\vspace*{-0.1cm}
\caption{ Higgs mass $m_H$ in the limits $\alpha =0 $ and $\alpha = 1$ obtained with Eqs.(\ref{eq410}) and (\ref{eq41}). $n_F^{(0)}$ is
the number of fermions (in the $SU(N)_{TC}$ fundamental representation) leading to an extreme walking gauge theory, while $n_F^{(1)}$ is just the number of fermions when TC is an ``scaled QCD" theory.}
\end{table}  

In the case of a large gauge group (for example, $SU(10)_{TC}$) the composite Higgs mass, in the extreme walking case,
can be almost as light as the present experimental limit. This result was also obtained by us in a much more involved
calculation of an effective potential for composite operators~\cite{we},  in this calculation the masses obtained 
are  of the order   shown in Table I (there are not appreciable differences for $\alpha =1$). However, in that calculation we 
take into account corrections due to contributions coming from top quark, if this effect is neglected, the mass values that we obtain
in Ref.\cite{we} are smaller than the ones shown in Table I.  Since the calculation presented in this work is much simpler than the one of Ref.\cite{we}, it
is important to know the origin of the differences.

The result of Elias and Scadron leading to Eq.(\ref{eq11}) was formulated by comparison of the homogeneous
BSE with the associated SDE, however the bound state properties are determined by the full BSE, which includes its normalization condition, as
clearly discussed by Llewellyn Smith~\cite{ls}. We credit the differences mentioned above to the fact that the BSE normalization condition changes the relation given by Eq.(\ref{eq11}) when the bound state wave function (that has the same formal expression of the self-energy) is characterized
by a hard asymptotic dynamics (the wave function decreases slowly as the momentum goes to infinity). In the sequence we just sketch a proof of such
result and the full calculation will be presented elsewhere~\cite{doff4}. The BSE normalization condition is given by~\cite{lane2}
\br
&2\imath&\!\!\!\left(\frac{F_\Pi}{m_{dyn}}\right)^2\!\!\! q_\mu = \imath^2\! \int d^4p \, 
Tr\left\{ {\cal P}(p)\left[\frac{\partial}{\partial q^\mu}F(p,q)\right] {\cal P}(p)\right\} \nonumber \\
&+& \int \, d^4pd^4k\, Tr\left\{{\cal P}(k)\left[\frac{\partial}{\partial q^\mu} K(p,k,q)\right]{\cal P}(p)\right\}\,\, ,
\label{eq43}
\er   
where 
\[
{\cal P}(p)  \equiv \frac{1}{(2\pi)^4}S(p)G(p)\gamma_5 S(p) \,\,,
\]
\[
F(p,q) =  S^{-1}(p+q) S^{-1}(p) \,\, ,
\]
$S(p)$ is the fermion propagator, $\Sigma (p)/m_{dyn}= G(p)$ and $K(p,k,q)$ is the BSE kernel.
Eq.(\ref{eq43}) can be written as
\[
2\imath \left({F_\Pi}/{m_{dyn}}\right)^2 q_\mu = I^0_\mu \, + \, I_\mu^K \,\, .
\]
Contracting the above equation with $q^\mu$ and computing it at $q^2=m_H^2$, after some algebra we
verify that the final equation can be put in the form $m_H^2 \, = \, 4m_{dyn}^2 (I^0 \, + I^K)$, where
$I^0$ and $I^K$ are the integrals of Eq.(\ref{eq43}) contracted with the momentum.

The simplest truncation
of the kernel $K(p,k;q)$ is the known rainbow-ladder approximation, where
\be
K^{rs}_{tu}(p,k;q\rightarrow 0) = - g^2 D_{\mu\nu}(k-p) \left( \gamma_\mu \frac{\lambda^a}{2} \right)_{tr} \left(\gamma_\nu \frac{\lambda^a}{2} \right)_{su} \,\, .
\label{eq51}
\ee
In this case $\partial_\mu^q K(p,k;q)\equiv 0$ and the second term of the normalization condition (Eq.(\ref{eq43})) does not contribute.
If we go beyond the rainbow-ladder approximation we obtain an expression of ${\cal{O}}(g^2(p^2))$ when compared to $I^0$. Neglecting $I^K$, introducing
Eq.(\ref{eq12}) into $I^0$, and considering the limit $\alpha \rightarrow 0$ we obtain
\be
m_{H_{BS}}^{(0)2} \approx 4 m_{dyn}^2 \left( \frac{1}{4}\frac{bg^2(m)(2\gamma - 1)}{(1 + \frac{bg^2(m)(2\gamma - 1)}{2})}\right) \,\,.
\label{eq511}
\ee
This expression clearly displays Lane's condition \cite{lane2}, i.e. $\gamma > 1/2$.
Therefore the normalization of the BSE introduce some modification in the scalar mass determination in the line proposed in
Ref.\cite{es,ds}. The $m_H^{(0)}$ values are decreased and turn out to be in agreement with Ref.\cite{we}, where the normalization of the composite field appears naturally. Of course, such modification is only appreciable in the limit where the
self-energy has a hard dynamics, i.e. in the walking (or conformal) limit of the theory. 

The complete expression for the composite scalar mass, in the extreme walking limit, comes from the junction of Eqs.(\ref{eq410}) and (\ref{eq511})
\be
m_{H_{BS}}^{(0)2}  \approx  m_{H}^{(0)2} \left( \frac{1}{4}\frac{bg^2(m)(2\gamma - 1)}{(1 + \frac{bg^2(m)(2\gamma - 1)}{2})}\right)  \,\, .
\label{eqf}
\ee
Table II shows the comparison of the same masses calculated in the Table I  with the effect of the 
BSE normalization for fermions in the fundamental representation of $SU(N)_{TC}$. Notice that the
normalization effect lower the masses to values that are not compatible with the present experimental limit on the
Higgs boson mass. The effect of the top quark mass (see Ref.\cite{we}) and the next order contribution to Eq.(\ref{eq43}) will
increase again the scalar mass, but it is quite probable that walking technicolor theories with fermions in the
fundamental representation may not provide good candidates for the symmetry breaking of the Standard Model. 

\begin{table}[t]
\begin{center}
\begin{tabular}{cccc}
 TC group  & $m_H^{(0)}$ & $m_{H_{BS}}^{(0)}$ &  $n_F^{(0)}$ \\ \\
\hline
   $SU(2)_{TC}$ &  585\,GeV    &  142\,GeV   &    8    \\ 
   $SU(3)_{TC}$ &  414\,GeV    &  106\,GeV   &    12    \\ 
   $SU(4)_{TC}$ &  304\,GeV    &   74\,GeV   &    14    \\   
\hline                   
\end{tabular}
\end{center}
\vspace*{-0.1cm}
\caption{ Comparison between the values of the mass obtained with Eq.(\ref{eq410}) ($m_H^{(0)}$) with the ones obtained after 
adding the effect of the  BSE normalization ($m_{H_{BS}}^{(0)}$) for fermions in the fundamental representation of $SU(N)_{TC}$. Notice that the
normalization effect lower the masses to values that are not compatible with the present experimental limit on the
Higgs boson mass.}
\end{table}

Besides the problem that we commented above, it can be noticed from Table I and II that we need a large number of fermions in order to obtain the extreme limit of walking
theories when the fermions are in the fundamental representation. A large number of fermions gives a too large $S$ parameter, whose
perturbative expression is
\be
S=\frac{1}{6\pi}\frac{n_f}{2}d(R) \,\, ,
\ee
where $d(R)$ is the dimension of the representation $R$. However, as shown by Dietrich and Sannino \cite{san}, we can
find theories with few fermions in higher representations which are compatible with the experimental data for $S$. 
A detailed calculation of the BSE normalization effect, including fermions in higher dimensional representations, for
different groups and with $S$ parameters consistent with the data is in preparation \cite{doff4}. The small mass values
of Table II appear as consequence of the number of fermions and group theoretical factors. It is possible to see
that the suppression in relation to the Fermi scale cannot be larger than a factor of ${\cal{O}}(10)$. This effect may 
even produce mass values that are not compatible with the actual experimental bound (as some values shown in Table II),
but examples of viable models will be presented in Ref.\cite{doff4}. Fortunately we can be confident that these masses
can be computed under certain controllable conditions, given the agreement between the BSE approach described here
and the effective action one of Ref.\cite{we}.

\par The measurement of a composite Higgs boson width will also be very important to determine the
underlying theory. In the Standard Model the computation of the width of the Higgs boson, with mass much greater
 than Z mass,  into $WW$ and $ZZ$ bosons can  be made as described in Ref.\cite{chivukula},
where it was determined that
$\Gamma_H  = \frac{3m^3_{H}}{32\pi v^2}$.
As discussed in Ref.\cite{schechter} it is well known that the Higgs sector of 
the standard electroweak model is identical to
the $SU(2)$ linear sigma model of mesons. The sigma corresponds to the Higgs boson while the
$\Pi^\pm$ and $\Pi^0$ appear eventually, in the Unitary gauge, as the longitudinal components of the
$W^\pm$ and $Z$ bosons. Although the QCD sigma and composite Higgs Lagrangian agree among themselves, it is clear that
one is not a simple scaled version of the other once the parameters, as the mass that we discussed
above, are different. However we can make use of the dynamically generated $SU(2)$ linear sigma
model to obtain the following result
\be
\Gamma_H \approx \frac{3m_{dyn}^3}{4F_{\Pi}^2} \,\,\, .
\label{eq50}
\ee
The Eq.(\ref{eq50}) is computed in terms of a triangle of fundamental fermions connecting
the $H$ boson to $\Pi$'s, in the same way that the width of the sigma meson is generated,
as discussed by Delbourgo and Scadron \cite{ds2}, in the limit $m_{\Pi} << m_H$. The fermionic
loop shrinks to the ``tree" level Lagrangian result but in terms of the appropriate parameters ($m_{dyn}$ 
and $F_{\Pi}$) \cite{ds2}. We can compute the scalar width of Eq.(\ref{eq50}) using the relation
between $m_{dyn}$ and $F_{\Pi}$ with $\Sigma (p)$ given by Eq.(\ref{eq12}), obtaining a result as a function of $m_H^{(\alpha)}$, which is 
given by the Eqs.(\ref{eq410}) and (\ref{eq41}) in the limits $\alpha =0$ or $\alpha =1$: 
\be
 \Gamma_H^{(\alpha)} \approx \frac{3n_{F}}{64}\frac{\left(m_H^{(\alpha)}\right)^3}{v^2} , \,\,\,\,\,\, (\alpha = 0 \,\, or \,\, 1) 
\label{eq51}
\ee
\begin{figure}[t]
\centering
\includegraphics[width=0.8\columnwidth]{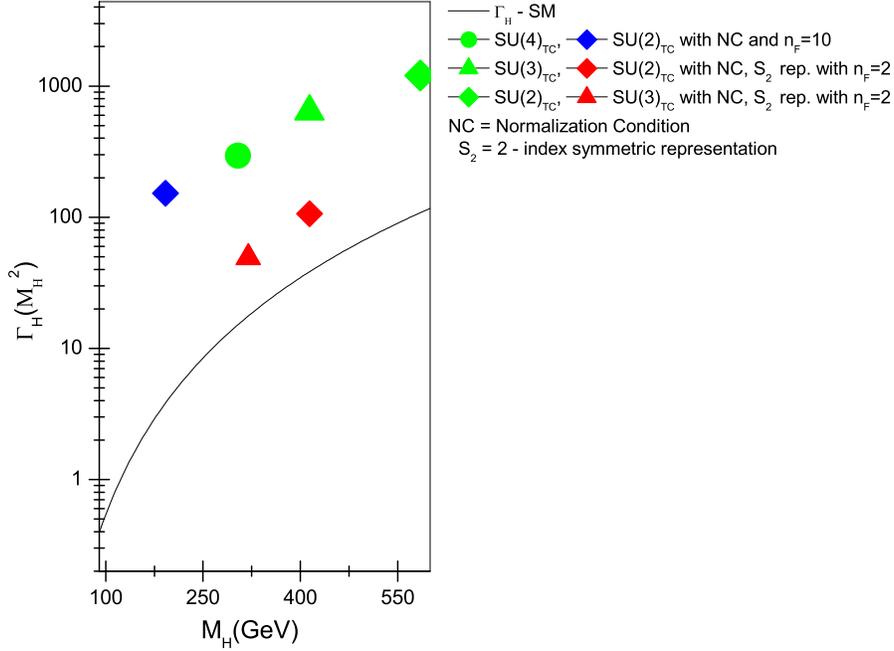}
\vspace*{-0.75cm}
\caption[dummy0]{ The figure shows the composite scalar width for some $TC$ groups ($SU(2)_{TC}$, $SU(3)_{TC}$ and $SU(4)_{TC}$ ) with fermions in the fundamental representation (green points) ($n_F = 8, \,\, 12, \,\, 14$,  respectively) as a function of its mass, in the limit $\alpha \rightarrow 0$, or the extreme walking limit. The solid line is the Standard Model (SM) result. In the same figure, we show the composite scalar masses corrected with the effect of the BSE normalization, (blue points) for fermions in the fundamental representation and ($n_F = 10$), and (red points) for fermions in the 2-index symmetric representation ($S_{2}$).}
\label{lamb6}
\end{figure}
\par In Fig.(\ref{lamb6}) we plot $\Gamma_H^{(0)}$ as a function of $m_H^{(0)}$, without and with
the effect of the BSE normalization condition, the solid line represents the width of the Standard Model Higgs boson. The points depicted in this figure represents the width for a composite Higgs boson when $\alpha \rightarrow 0$. The different TC points
lie above the SM curve, but follow similar behavior due to the $SU(2)$ symmetry leading to Eq.(\ref{eq51}). The large $n_{F}$ values for some of the points of Fig.(1) may lead to problems with high precision electroweak measurements, but these will eventually be overcome in viable TC models \cite{hs}. For this reason we have already included in Fig.(\ref{lamb6}) some points determined for $SU(2)_{TC}$ and $SU(3)_{TC}$ theories with technifermions in
the 2-index symmetric representation, which imply in smaller $n_F$ and $S$ values \cite{doff4}. The important fact is that the $(\Gamma_H \times m_H)$ curve may give hints about the TC gauge theory.  In the limit $\alpha \rightarrow 1$ we obtain the early results of TC as an scaled QCD version, although this case is not phenomenologically interesting we verified that the composite scalar mass is larger as well as its width, the curve $(\Gamma_H \times m_H)$ is flatter and would not be useful to discriminate a TC theory.

We calculated the mass and width of a composite scalar Higgs boson. If a composite scalar is found at LHC 
our result show how its mass and width will give some information about the underlying strongly interacting theory. Here, as well as in
Ref.\cite{we}, we verified that we may have a light composite Higgs boson in the extreme walking regime. There are
some limitations in our calculation: A composite scalar boson could mix with other scalars, formed, for instance,
by technigluons, which is a problem already discussed for the sigma meson in QCD~\cite{ds}, but not taken into account here. We have not considered the contribution of a heavy top quark for the scalar mass (this possible contribution was discussed in Ref.\cite{we}). Techniquarks may have a current mass and will also introduce an extra contribution to the SDE solution and modify our prediction. In a viable TC model the technipions are not massless and their mass will change the calculation of the width. Some of these problems will be discussed elsewhere, nevertheless all these effects should not change drastically the mass and width values that we discussed here, in such a way that their measurement provide a sound hint  of the gauge theory that form such bound state. Finally, it is quite interesting that the scalar composite masses can be computed under certain controllable approximations, as in the Bethe-Salpeter approach, and the results shown here confirm the ones obtained in a more complicated calculation
as the one of the effective action of Ref.\cite{we}, and the importance of considering the BSE normalization condition in walking gauge theories.

\section*{Acknowledgments}
We thank P. S. Rodrigues da Silva and A. C. Aguilar for useful discussions. This research was partially supported by the Conselho Nacional de Desenvolvimento Cient\'{\i}fico e Tecnol\'ogico (CNPq).


\begin {thebibliography}{99}

\bibitem{nl} Y. Nambu and G. Jona-Lasinio, {\it Phys. Rev.} {\bf 122}, 345  (1961).
\bibitem{ds} R. Delbourgo and M. D. Scadron, {\it Phys. Rev. Lett.} {\bf 48}, 379 (1982).
\bibitem{polosa} N. A. Tornqvist and M. Roos, {\it Phys. Rev. Lett.} {\bf 76}, 1575 (1996);
N. A. Tornqvist and A. D. Polosa, {\it Nucl.Phys. A} {\bf 692}, 259 (2001); {\it Frascati Phys.Ser.} {\bf 20}, 385 (2000);
A. D. Polosa, N. A. Tornqvist, M. D. Scadron and V. Elias, {\it Mod. Phys. Lett. A} {\bf 17}, 569 (2002).  
\bibitem{tc} S. Weinberg, {\it Phys. Rev. D} {\bf 19}, 1277 (1979); L. Susskind, {\it Phys. Rev. D} {\bf 20}, 2619 (1979).
\bibitem{hs} F. Sannino, hep-ph/0809.0793; idem hep-ph/0812.1788, A. Martin, hep-ph/0812.1841 and references therein.
\bibitem{we} A. Doff, A. A. Natale and P. S. Rodrigues da Silva, {\it Phys. Rev. D} {\bf 77}, 075012 (2008).
\bibitem{pdg} The Review of Particle Physics, W.-M. Yao {\it et al.}, {\it J. of Phys. G} {\bf 33}, 1 (2006).
\bibitem{es} V. Elias and M. D. Scadron, {\it Phys. Rev. Lett.} {\bf 53}, 1129 (1984).
\bibitem{san} F. Sannino, {\it Int. J. Mod. Phys. A} {\bf 20}, 6133 (2005); D. D. Dietrich, F. Sannino and K. Tuominen, {\it Phys. Rev. D} {\bf 72}, 055001 (2005);  N. Evans and F. Sannino, hep-ph/0512080; D. D. Dietrich, F. Sannino and K. Tuominen, {\it Phys. Rev. D} {\bf 73}, 037701 (2006); D. D. Dietrich and F. Sannino, {\it Phys. Rev. D} {\bf 75}, 085018 (2007); R. Foadi, M. T. Frandsen, T. A. Ryttov and F. Sannino, {\it Phys. Rev. D} {\bf 76}, 055005 (2007); R. Foadi, M. T. Frandsen and F. Sannino, hep-ph/0712.1948. 
\bibitem{walk} B. Holdom, {\it Phys. Rev.} {\bf D24},1441 (1981);{\it Phys. Lett.}
{\bf B150}, 301 (1985); T. Appelquist, D. Karabali and L. C. R.
Wijewardhana, {\it Phys. Rev. Lett.} {\bf 57}, 957 (1986); T. Appelquist and
L. C. R. Wijewardhana, {\it Phys. Rev.} {\bf D36}, 568 (1987); K. Yamawaki, M.
Bando and K.I. Matumoto, {\it Phys. Rev. Lett.} {\bf 56}, 1335 (1986); T. Akiba
and T. Yanagida, {\it Phys. Lett.} {\bf B169}, 432 (1986).
\bibitem{frere} J.-M. Fr\`ere, {\it Phys. Rev. D} {\bf 35}, 2625 (1987).
\bibitem{doff1}A. Doff and A. A. Natale, {\it Phys. Lett. } {\bf B537}, 275 (2002).
\bibitem{doff2} A. Doff and A. A. Natale, {\it Phys. Rev.} {\bf D68}, 077702 (2003).
\bibitem{soni} J. Carpenter, R. Norton, S. Siegemund-Broka and A. Soni,  {\it Phys. Rev. Lett.} {\bf 65}, 153 (1990).
\bibitem{soni2} J. D. Carpenter, R. E. Norton and A. Soni, {\it Phys. Lett.} {\bf B 212}, 63 (1988).
\bibitem{lane2} K. Lane, {\it Phys. Rev.} {\bf D10}, 2605 (1974).
\bibitem{pagels} H. Pagels and S. Stokar, {\it Phys. Rev.} {\bf D20}, 2947 (1979).
\bibitem{doff3} A. Doff and A. A. Natale, {\it Phys. Lett.} {\bf B641}, 198 (2006).
\bibitem{ls} C. H. Llewellyn Smith, {\it Il Nuovo Cimento} {\bf A60}, 348 (1969).
\bibitem{doff4} A. Doff, A. A. Natale and P. S. Rodrigues da Silva, (in preparation).
\bibitem{chivukula} R. Sekhar Chivukula, Michael J. Dugan and  Mitchell Golden,  {\it Phys. Lett.} {\bf B336}, 62 (1994).
\bibitem{schechter} A. A.-Rehim, D. Black, A. H. Fariborz, S. Nasri and J. Schechter, {\it Phys. Rev.} {\bf D68}, 013008 (2003).
\bibitem{ds2} R. Delbourgo and M. D. Scadron, {\it Mod. Phys. Lett. A} {\bf 10}, 251 (1995).



\end {thebibliography}
\end{document}